\documentclass[preprint,preprintnumbers,nofootinbib]{revtex4-1} 
\usepackage{graphicx, color} 
\usepackage{amsmath,amssymb}
\usepackage{ulem} 
\usepackage{epsf}
\usepackage{color}

\usepackage[italicdiff]{physics}
\usepackage{bm}
\usepackage{mathtools}
\usepackage{enumitem}
\usepackage{wrapfig}
\usepackage{footnote}
\usepackage{here, comment}
\usepackage{bbm}
\usepackage{shorttoc}

\usepackage{subfig}
\usepackage{here, comment}
\usepackage{tikz-feynman}
\usepackage{tikz-feynhand}

\allowdisplaybreaks[1]

\newcommand{\nn}{\nonumber\\}

\newcommand{\be}{\begin{equation}}
\newcommand{\e}{\end{equation}}
\newcommand{\aln}[1]{\begin{align}#1\end{align}}
\newcommand{\ga}[1]{\begin{gather}#1\end{gather}}

\begin{document}
	\pagestyle{empty}

	\preprint{KEK-TH-2323}
	
	\title{Wilsonian Effective Action and  Entanglement Entropy
	} 
	\author{Satoshi Iso$^{a,b}$} 
	\email{satoshi.iso@kek.jp}
	\author{Takato Mori$^{a,b}$}
	\email{moritaka@post.kek.jp}
	\author{Katsuta Sakai$^a$}
	\email{sakaika@post.kek.jp}
	
	\affiliation{
		$^a$ KEK Theory Center, High Energy Accelerator Research Organization (KEK), Oho 1-1, Tsukuba, Ibaraki 305-0801, Japan.\\
		$^b$ The Graduate University for Advanced Studies (SOKENDAI), Oho 1-1, Tsukuba, Ibaraki 305-0801, Japan.
	}
	
	\begin{abstract} 
This is a continuation of our previous works on entanglement entropy (EE)  in interacting field theories. 
In  \cite{Iso:2021vrk}, we have proposed the notion of 
$\mathbb{Z}_M$ gauge theory on Feynman diagrams to calculate EE in quantum field theories and 
shown that EE consists of two particular contributions from propagators and vertices. 
As shown in the next paper \cite{Iso:2021rop}, 
the purely non-Gaussian contributions from interaction vertices 
can be interpreted as renormalized correlation functions of composite operators. 
In this paper, we will first 
provide a unified matrix form of EE containing both contributions from propagators and (classical) vertices, 
and then extract further non-Gaussian contributions  based on the framework of the 
 Wilsonian renormalization group. 
It is conjectured that the  EE in the infrared  is given by a sum of all the vertex contributions in 
the Wilsonian effective action.

	\end{abstract}
	\maketitle
	\pagestyle{plain}
\section{Introduction}
\label{s:intro}
Entanglement entropy (EE) captures correlations 
for bipartite entangled states between two subspaces, in particular spatially separated regions, 
and has been widely investigated in conformal field theories (CFTs)~\cite{Calabrese:2004eu,Calabrese:2009qy,Ruggiero:2018hyl,Hung:2014npa,Casini:2010kt}, 
perturbations from CFTs~\cite{Rosenhaus:2014zza,Rosenhaus:2014woa,Rosenhaus:2014ula} or
in the context of AdS/CFT correspondence~\cite{Ryu:2006bv,Ryu:2006ef,Hubeny:2007xt, Nishioka_2009}
(for review, \cite{Nishioka_2009, RevModPhys.90.035007}).
In free quantum field theories (QFTs), where 
the vacuum wavefunctional is Gaussian, EE is well-understood~\cite{Calabrese:2004eu,Casini_2009} and we can perform  explicit calculations~\cite{PhysRevA.70.052329,Katsinis:2017qzh,Bianchi:2019pvv,Lewkowycz:2012qr, Herzog:2013py}. 
On the other hand, 
we have little understanding of EE in general interacting QFTs, 
apart from exactly solvable cases \cite{Donnelly:2019zde} or some supersymmetric theories~\cite{Nishioka_2009,Jafferis:2011zi,Pufu:2016zxm,Nishioka:2013haa,RevModPhys.90.035007}. 
EE in interacting theories are 
discussed in perturbative~\cite{Hertzberg:2012mn,Chen:2020ild}, nonperturbative~\cite{PhysRevB.80.115122,Akers:2015bgh,Cotler:2015zda,Fernandez-Melgarejo:2020utg,Fernandez-Melgarejo:2021ymz,Whitsitt:2016irx,Hampapura:2018uho},  
lattice~\cite{Wang_2014,Buividovich:2008kq,Buividovich:2008gq,Itou:2015cyu,Rabenstein:2018bri},  
or in terms of variational trial wave functions~\cite{Cotler:2015zda,Fernandez-Melgarejo:2020utg,Fernandez-Melgarejo:2021ymz}.

In interacting QFTs, EE  is  divergent and needs regularizations. 
First,  EE is UV divergent because there are infinitely many degrees of freedom. This divergence occurs
even in a free theory. Second, in interacting theories, the infinite degrees of freedom cause 
 UV divergences in physical parameters and renormalizations are necessary to extract finite results.
Finally, if a theory contains massless fields, it may cause IR divergence, but 
in this paper, we consider a massive theory so that  the IR divergences are assumed to be absent. 

The EE in the infrared limit, or its variation with respect to some parameters such as mass or coupling constants, 
is determined by correlations of the renormalized vacuum wave function
and should be independent of the UV cutoff. 
In this sense, the IR part of EE must be determined by the IR fixed point of the renormalization group (RG).
On the other hand, EE at a fixed scale
will change along the RG flow on which a theory transmutes 
from one fixed point to another. In order to 
understand these behaviors of EE, Wilsonian approach of renormalization \cite{PhysRevB.4.3174, PhysRevB.4.3184,Wilson:1973jj,Polchinski:1983gv} will be useful. 
The Wilsonian effective action (EA) describes a flow of effective actions at a given scale
where all higher momentum fluctuations of fields are integrated out, 
together with rescaling of  the momentum 
so as to make the UV cutoff back to the original one. 

In this paper, we will investigate  EE in interacting field theories based on the Wilsonian
picture of renormalization combined with our previous works  \cite{Iso:2021vrk,Iso:2021rop}
based on the $\mathbb{Z}_M$ gauge theory on Feynman diagrams.
In \cite{Iso:2021vrk}, we succeeded to extract two particular contributions to EE in interacting 
field theories. One is the Gaussian contributions written in terms of 
renormalized two-point correlation functions in the
two-particle irreducible (2PI) formalism. 
Another set of important contributions comes from {\it classical} vertices, which reflects
non-Gaussianity  of the vacuum wave function.
In \cite{Iso:2021rop}, we showed that the vertex contributions 
 can be interpreted as contributions from 
renormalized two-point correlation functions of {\it composite operators.} 
These results are obtained by evaluating EE in the notion of  the $\mathbb{Z}_M$ gauge theory on Feynman diagrams, 
whose picture is derived from the replica method of EE (the number of replicas $n$ is replaced by   $n=1/M$). 
In the formulation, EE is given by a sum of various configurations of $\mathbb{Z}_M$ fluxes on 
plaquettes in Feynman diagrams and the above two  contributions to EE are 
described by two particular types of flux configurations. Thus, an important question left unanswered 
is how to extract other contributions described by other configurations of fluxes. 
In this paper, we address this question in the framework of the Wilsonian RG,
where a variety of {\it quantum} vertices appears as the energy scale decreases. 

For this purpose, we first generalize our previous results to 
describe operator mixings. 
We also give a natural, unified description of the contributions to EE
from propagators and vertices in 
a matrix form. 
This unified description is one of the main results of the present paper. 
By using this generalized expression of EE, we conjecture that the IR  part of EE
is given by a sum of the {propagator and} vertex contributions in the Wilsonian EA.

The paper is organized as follows. 
In Sec.\ref{s:summary}, we first briefly summarize the notion of the  
$\mathbb{Z}_M$ gauge theory on Feynman diagrams, two particular contributions to EE from 
propagators and vertices in the $\phi^4$ scalar theory, and an 
interpretation of the vertex contribution  in terms of a correlator of a composite operator. 
In addition, we give a unified description of both contributions in a matrix form. 
In Sec.\ref{s:vertex}, we generalize it when various operators are mixed with each other and also when
the composite  operators have spins in the two-dimensional spacetime normal to the boundary. 
In Sec.\ref{Wilson}, we discuss the IR behavior of EE in the framework of the  Wilsonian RG and
give a conjecture that EE {in the IR} is given by a sum of the propagator and vertex contributions in the Wilsonian EA. 
Finally, we give  conclusions   in Section \ref{s:discussion}.
{In Appendix \ref{appenarea}, we prove the area law for R\'{e}nyi entropy and
the capacity of entanglement~\cite{PhysRevLett.105.080501,deBoer:2018mzv}.
In Appendix \ref{appencomp}, we give a proof that all the single twist contributions from  vertices are written 
in  the 1-loop type expression of composite operators. This is a generalization of the proof for the 
propagator contributions based on the 2PI formalism.
}

 \section{Summary of Previous Works}
\label{s:summary}
In this section, we first summarize our previous works in \cite{Iso:2021vrk} and \cite{Iso:2021rop}, 
and then introduce a new concept of the {\it generalized} 1PI in order to unify the contributions to EE
from propagators and vertices. 
\subsection{$\mathbb{Z}_M$ gauge theory on Feynman diagrams}
\label{s:ZM}
\begin{figure}[t]
	\begin{tabular}{c}
		\hspace*{-0.05\linewidth}
		\begin{minipage}{0.35\hsize}
			\centering
			\includegraphics[width=\linewidth]{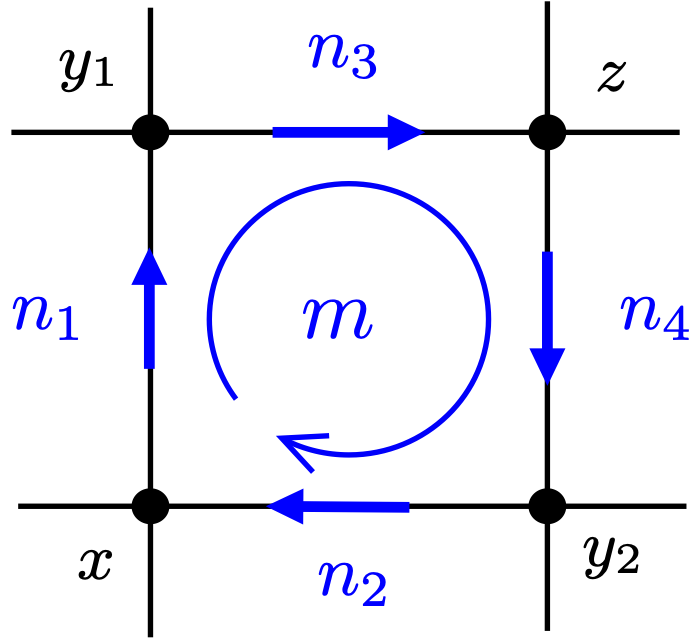}
		\end{minipage}
	\end{tabular}
\caption{
The  figure shows a $\mathbb{Z}_M$ twist configuration on a Feynman diagram.
{Given} twists  $\{n_i\}$ 
assigned on   propagators, 
a flux of twists {$m=0,\cdots M-1$} is defined on a plaquette  by 
a sum of twists around the plaquette $m=\sum_i n_i \mod M$. 
The flux is invariant under  $\mathbb{Z}_M$ gauge transformations on vertices.
}
\label{Fig1}
\end{figure}

Entanglement entropy of a subsystem $A$ is defined by 
\aln{ S_{EE}=-\Tr_{A}\rho_{A}\log\rho_{A},
}
where $\rho_A=\Tr_{\bar{A}}\rho_\mathrm{tot}$ is a reduced density matrix of $\rho_{\mathrm{tot}}$ obtained
by integrating out the complementary system $\bar{A}$ in the Hilbert space. 
In this paper, we take $A$ as a half space on a time slice in a $(d+1)$-dimensional spacetime 
and utilize the  orbifold method~\cite{Nishioka_2007,He:2014gva
} 
to calculate $S_{EE}$. 
This method is a variation of the replica trick for EE {in which EE is given through the $n\rightarrow 1$ limit of R\'{e}nyi entropy $S_n=\frac{1}{1-n} \log \Tr \rho_A^n$}~\cite{Calabrese:2004eu}; {by taking the replica parameter $n=1/M$,} we consider free energy of interacting quantum field theories on 
the orbifold $\mathbb{R}^2/\mathbb{Z}_M\times \mathbb{R}^{d-1}$ {denoted by $F^{(M)}$}. Then, EE is written as 
\aln{
	S_{EE}=-\frac{\partial \left(M F^{(M)}\right) }{\partial M}\bigg|_{M\to 1}.
	\label{eq:EE_M}
}
Since a physical state of the $\mathbb{Z}_M$ orbifold theory is invariant under the $\mathbb{Z}_M$ 
projection operator, 
\aln{
\hat{P}=\sum_{n=0}^{M-1}\frac{ \hat{g}^{\, n} }{M},
}
where 
$\hat{g}$ is a $2\pi/M$ rotation operator around the origin,
the orbifold theory can be interpreted as the $\mathbb{Z}_M$ gauge theory on Feynman diagrams. 
Namely, each propagator in a Feynman diagram is sandwiched by the projection operators; this corresponds to assigning a \textit{twist} $n_i$ on the $i$-th propagator $G(\hat{g}^{n_i}x,y)$ and summing over all independent configurations of such twists. 
Then, the notion of $\mathbb{Z}_M$ gauge symmetry appears since we can rotate away 
some of the twists of propagators by the $\mathbb{Z}_M$ gauge transformations on vertices of the Feynman diagram.
As a result, we can classify
independent configurations of twists
up to  $\mathbb{Z}_M$ transformations in terms of 
$\mathbb{Z}_M$ fluxes of twists on plaquettes. 
Here a flux of twists on a plaquette is defined by 
a sum of twists around the plaquette as shown  in Fig.\ref{Fig1}.
\begin{figure}[t]
\centering
\includegraphics[width=0.25\linewidth]{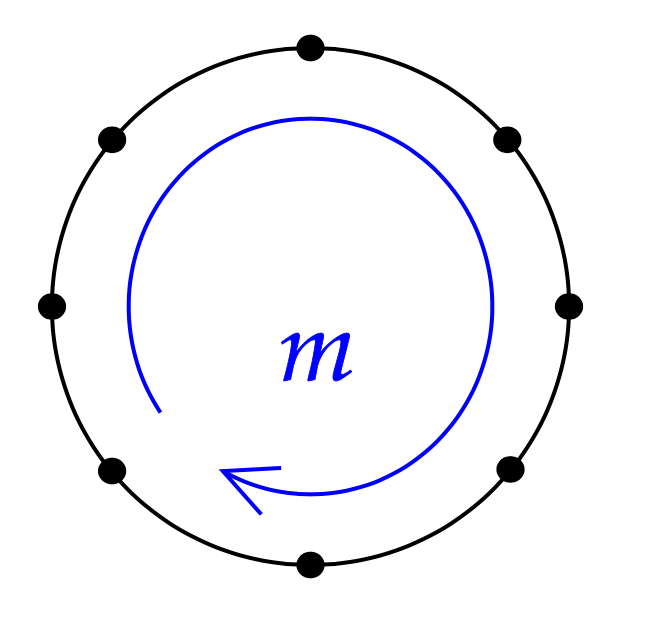}
\caption{
There is a single twist for a 1-loop diagram.
The number of independent twists is unchanged
even if we divide  the propagator into multiple connecting pieces as drawn in the figure.
}
\label{Fig2}
\end{figure}
A simple example of the $\mathbb{Z}_M$ invariant configuration is given by Fig.\ref{Fig2}. 
The 1-loop diagram has only one plaquette and the $\mathbb{Z}_M$ invariant twist is given by the flux {$m$}.
The number of independent flux is always 1 even if we divide the propagator into several
connecting pieces. 
One may think that each propagator can be twisted separately, but 
such multiplicities are removed by the $\mathbb{Z}_M$ transformations on vertices
connecting the divided propagators. Thus, there is only one independent twist in the 1-loop diagram.
This property is essential to prove our main result of Eq.(\ref{EE-generalform2}) and
responsible for the fact that
EE can be written as a sum of 1-loop type diagrams of various composite operators. See the proof in Apendix \ref{appencomp}.
See also the discussions  in Sec.IV.B and Sec.IV.C 
in our previous paper \cite{Iso:2021rop}.

\subsection{Propagator contributions to EE}
\label{s:propagatorEE}
In order to evaluate EE in Eq.(\ref{eq:EE_M}), we need to extract all the configurations
of fluxes that do not vanish in the $M \rightarrow 1$ limit.  
In the previous papers, we have shown that, if all the fluxes of twists are zero, they  
do not contribute to EE in Eq.(\ref{eq:EE_M}), which assures the area law of EE\footnote{It is easy to see that this property  also holds  for the R\'{e}nyi entropy or entanglement capacity, and so does the area law. For more details, see Appendix \ref{appenarea}}. 
Among various configurations  contributing to EE, we have focused on two particular 
configurations. 
\begin{figure}[t]
\centering
\includegraphics[width=9cm]{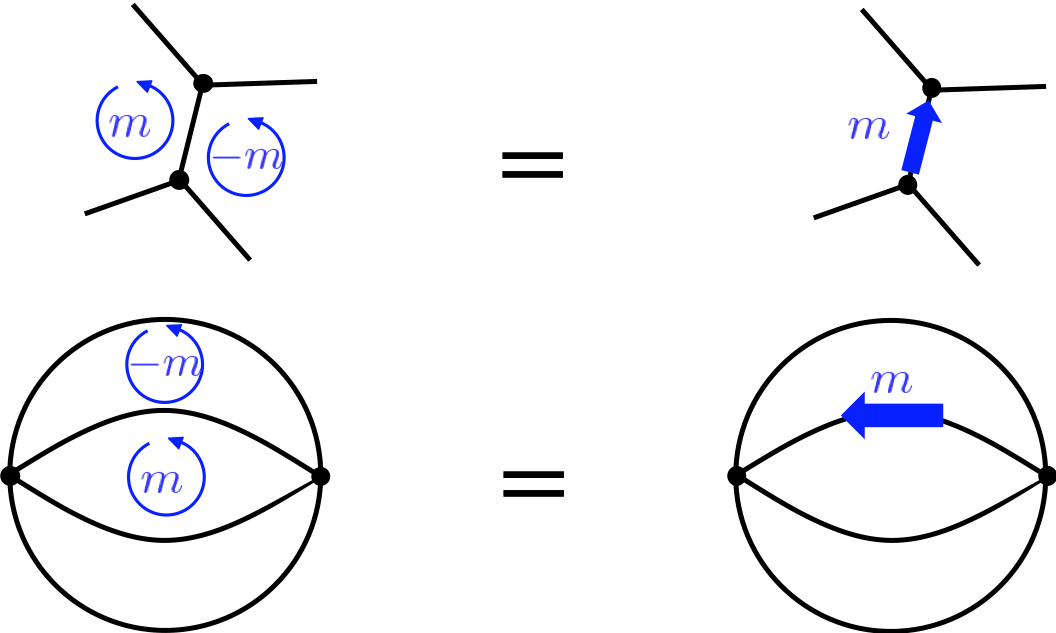}
\caption{
Twist of a propagator: 
if the fluxes of plaquettes straddling a shared propagator are given by $m$ and $-m$, 
such a configuration is interpreted  as a  twist of the shared 
propagator.   
The upper figures show a relevant part  in a general Feynman diagram that twists the propagator.
The lower figures are an example of such a configuration in a 3-loop diagram.
}
\label{f:proptwist}
\end{figure}
The first type of configurations are given in 
Fig.\ref{f:proptwist}
 and can be interpreted as a twist of the propagator. 
The simplest configuration of this type is given by Fig.\ref{Fig2} and 
already present in a free field theory.  
Interactions induce renormalization of physical quantities, such as a mass, appearing in the EE. 
A seminal calculation was studied by Hertzberg \cite{Hertzberg:2012mn}
and completed in our previous papers 
based on the two-particle irreducible formalism (2PI) where we have shown that 
the propagator contributions are exactly given by
\aln{
S_{\text{propagator}} & =- \frac{V_{d-1} }{6} 
 \int^{1/\epsilon} \frac{d^{d-1}k_{\parallel}} {2(2 \pi)^{d-1}}  
  \log \left[ G^{-1}(\bm{0}; k_\parallel) \epsilon^2 \right] 
  \label{e:EEprop}    \\
  &  =-  \frac{V_{d-1} }{6} 
 \int^{1/\epsilon} \frac{d^{d-1}k_{\parallel}} {2(2 \pi)^{d-1}}  
  \log \left[  ( G_0^{-1}  - \Sigma) 
  (\bm{0}; k_\parallel) \epsilon^2 \right], 
\label{e:EE2PIprop}
}
where $G$ is the renormalized propagator\footnote{The same symbol $G$ is used
to represent the Green function in coordinate and momentum spaces for notational simplicity. 
They are distinguished by their  arguments if necessary.} 
and $V_{d-1}$ is the volume of the boundary $\partial A$.
The $(d+1)$-dimensional  momentum is written as $({\bm k}, k_\parallel)$, where ${\bm k}$ is the
two-dimensional components of time and the direction normal to the boundary and $k_\parallel$
is the $(d-1)$-dimensional components parallel to the boundary. 
The UV cutoff $\epsilon$ is introduced.  
Writing the full inverse propagator  as $G^{-1}=(G_0^{-1}  - \Sigma) $, 
the logarithm can be expanded as a sum,
\aln{
-\log G^{-1} = -{\log}G_0^{-1} + \sum_{n=1} \frac{(G_0 \Sigma)^n}{n}.
} 
Then, it can be interpreted as a chain of free propagators connected by the self-energy $\Sigma$. 
On the other hand, the full propagator $G$ itself is expanded similarly, but without the $1/n$ factor.
This $1/n$ factor in the EE 
comes from the redundancy of twists:  twisting $n$ propagators in the chain
is not independent.  
There is only a single twist in the plaquette as explained at the end of Sec. \ref{s:ZM}.
If we twisted every propagator in the chain, it would overcount the contributions to EE from the 
1-loop Feynman diagram.

\subsection{Vertex contributions to EE and generalized 1PI}
\label{s:vertexEE}
\begin{figure}[t]
	\centering
	\includegraphics[width=10cm]{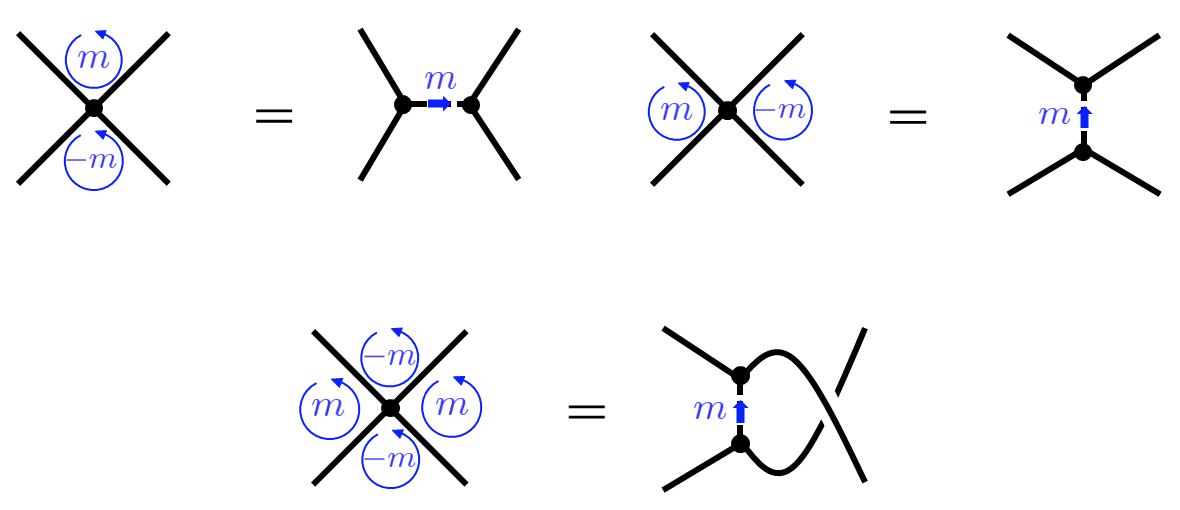}
	\caption{
Twist of a vertex: three types of 
configurations can be attributed to a twist of the vertex.
The dotted lines in the figures on the right-hand sides represent a 
{virtual} propagator that appears by opening the four-point vertex into a pair of three-point vertices
by a delta function. 
The twist of a vertex is interpreted as a twist of the dotted propagator. 
	}
	\label{f:verttwist}
\end{figure}
\begin{figure}[t]
\centering
\includegraphics[width=17cm]{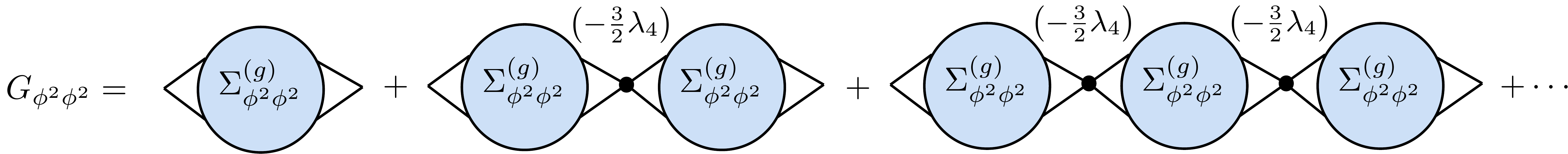}
\caption{
A Green function of a composite operator can be written in terms of the generalized self-energy $\Sigma^{(g)}_{\phi^2\phi^2}$,
which is 1PI with respect to the propagator of the composite operator at the vertex as well as the fundamental field. 
}
\label{f:G-1PI}
\end{figure}
\begin{figure}[t]
\centering
\includegraphics[width=9cm]{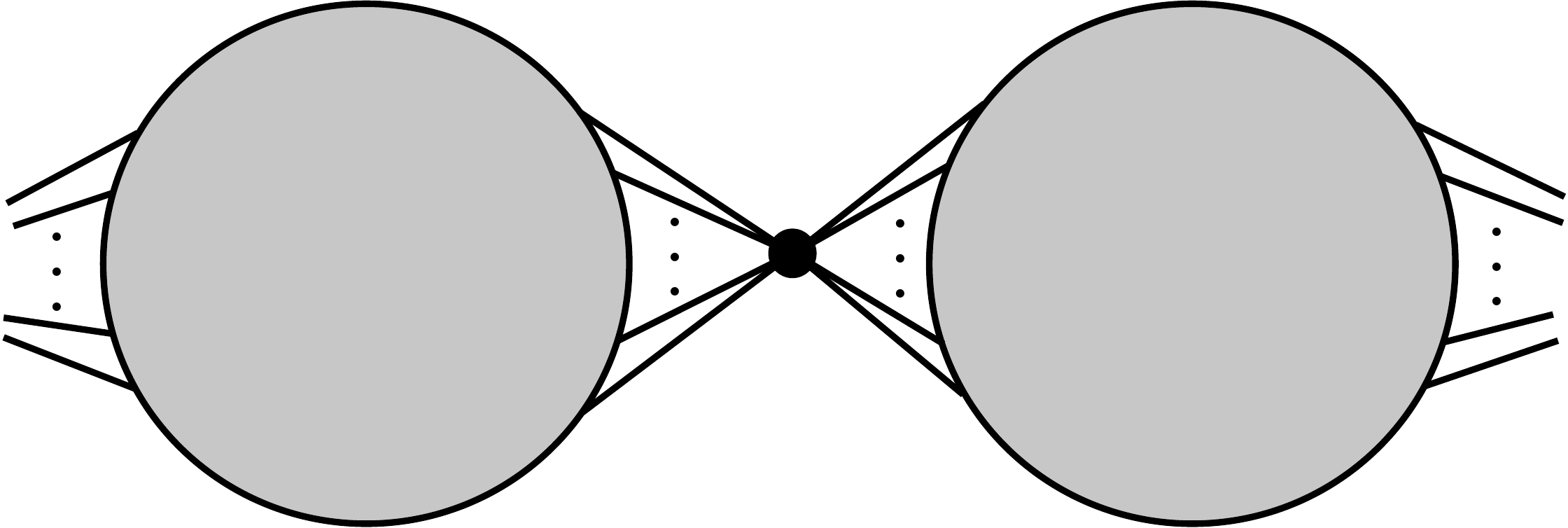}
\caption{
Beads diagram: this diagram is NOT  1PI in the {\it generalized} sense 
since it is separable by cutting the  propagator at the opened vertex.
}
\label{f:g1PI}
\end{figure}

The second type of  configurations of fluxes we are going to focus on 
is given by Fig.\ref{f:verttwist} for the $\phi^4$ scalar theory, {$\mathcal{L}_{\mathrm{pot}}=\lambda_4 \phi^4 /4$}.
These configurations of fluxes are
 interpreted as twists of the interaction vertices, which in turn, regarded as twists of the corresponding composite operators.
 This interpretation is obtained by 
opening a 4-point vertex into two 3-point vertices and assign the twist to the propagator
connecting the two 3-point vertices. 
Corresponding to three different channels of {the opening,} $s$, $t$, and $u$, 
there are three different configurations of fluxes and contributions to EE respectively.
If  different quantum numbers 
are assigned to these
three channels, we can utilize a method of auxiliary fields
and EE is given by a {\it sum} of propagator contributions of three different auxiliary fields as shown in \cite{Iso:2021rop}. 
If composite operators propagating three channels are mixed like the $\phi^4$-theory, 
we cannot use the method of auxiliary fields, but from diagrammatic analysis (see Appendix \ref{appencomp}), we can show that
it is written in terms of   a correlation function of composite operators. 
In the case of the $\phi^4$-theory, it is given~\cite{Iso:2021rop} by
\aln{
		S_{\text{vertex}}= 
		\frac{V_{d-1}}{6}
		\int^{1/\epsilon}\frac{d^{d-1}k_\parallel}{2(2\pi)^{d-1}}\mathrm{log}\,\left[1 - \frac{3}{2}\lambda_4 \,
		G_{\phi^2 \phi^2}(\bm{0},k_\parallel)
		\right] ,
		\label{e:EEcomp}
	}
where
	\aln{
		G_{\phi^2\phi^2}(x-y):= & \left<\,[\phi^2](x)\, [\phi^2](y)\,\right> .
	}
In the following, the cutoff $\epsilon$ is not explicitly written for notational simplicity, 
as it can be recovered by dimensional analysis. 
The square brackets  $[{\cal O}]$ represent the normal ordering of an operator ${\cal O}$. 
The coefficient $-3\lambda_4 /2$ is a product of  $-\lambda_4 /4$ 
and $6$, 
where $-\lambda_4/4$ is the coefficient in front of the interaction vertex 
and the coefficient $6$ is a combinatorial factor for separating four $\phi(x)$'s into 
{a pair of $\phi^2(x)$ and $\phi^2(y)$.} 
As shown in Fig.\ref{f:G-1PI}, 
the Green function of the composite operator can be written as
\aln{
G_{\phi^2\phi^2} = \,&\Sigma^{(g)}_{\phi^2\phi^2}
+\left(\frac{-3\lambda_4 }{2} \right)(\Sigma^{(g)}_{\phi^2\phi^2})^2 + 
\left(\frac{-3\lambda_4 }{2} \right)^2 (\Sigma^{(g)}_{\phi^2\phi^2})^3 + \cdots
\nn
=\, & \frac{\Sigma^{(g)}_{\phi^2\phi^2}} {1-  \left(\frac{-3\lambda_4 }{2} \right) \Sigma^{(g)}_{\phi^2\phi^2}},
\label{G=sumofg1PI}
}
where $\Sigma^{(g)}_{\phi^2\phi^2}$ is the 1PI self-energy of $[\phi^2]$  
{in a \textit{generalized} sense.} We call it g-1PI. 
Namely,  the quantity with the superscript $(g)$ 
 does not contain a diagram like Fig.\ref{f:g1PI}
that is separable by cutting a vertex in the middle. 
We call such a diagram a beads diagram:
1PI in the ordinary sense but not in the generalized sense. Thus these beads diagrams 
are not included in g-1PI diagrams. 

By using Eq.(\ref{G=sumofg1PI}),
we can rewrite Eq.(\ref{e:EEcomp}) as
\aln{
S_{\text{vertex}}= 
		 - \frac{V_{d-1}}{6}
		\int^{1/\epsilon}\frac{d^{d-1}k_\parallel}{2(2\pi)^{d-1}}\mathrm{log}\,\left[1 -  \left( -\frac{3}{2}\lambda_4 \right) \,
		 \Sigma^{(g)}_{\phi^2\phi^2}
		\right] .
		\label{e:EEcomp2}
}
In the following equations including Eq.(10), the argument $(\bm{k}=\bm{0},k_\parallel)$ of the integrand for the $k_\parallel$ integral is implicit.
Now we can write 
both the propagator and vertex contributions in Eq.(\ref{e:EE2PIprop}) and Eq.(\ref{e:EEcomp2})
in a unified matrix form as
\aln{
S_{EE}^{(\phi^4)} = - \frac{V_{d-1}}{6}
		\int^{1/\epsilon}\frac{d^{d-1}k_\parallel}{2(2\pi)^{d-1}}\mathrm{tr}\,\mathrm{log}
	\left[\hat{G}_0^{-1} - \hat\lambda \hat{\Sigma}^{(g)} \right],
	\label{eq:EE_phi4}
}
where
\aln{
  \hat{G}_0 =&   \left( \begin{array}{cc} G_{0} & 0 \\0 & 
  1  \end{array}\right)  , 
  \   \ 
  \hat\lambda = \left(\begin{array}{cc}1  & 0   \\ 0  & -3 \lambda_4/ 2 \end{array}\right) ,   
  \   \
 \hat{\Sigma}^{(g)} =    \left(\begin{array}{cc}  \Sigma^{(g)}    &{0}   
  \\ 0  & \Sigma^{(g)}_{\phi^2\phi^2}\end{array}\right) .
  \label{GLS1}
 }

In the following, we first generalize these results to include higher-point vertices whose composite operators 
are mixed 
in a complicated way. Then, we apply the concept of Wilsonian effective action to extract further contributions 
to the IR part of the EE. 
It is important to note that the form of Eq.(\ref{eq:EE_phi4}) 
 is convenient for a unified description in the following discussions, but it is 
always possible to go back to the form like Eq.(\ref{e:EEcomp}), where 
the vertex contributions are written  in terms of the ordinary renormalized propagators without the superscript $(g)$. 
Also, note that all the single twist contributions from a vertex can be written in the above 1-loop type formula,
 Eq.(\ref{e:EEcomp}) or Eq.(\ref{GLS1}). 
In the case of the propagator contributions to EE, we have proved the statement by using the 2PI formalism in 
\cite{Iso:2021vrk,Iso:2021rop}. Here we use a diagrammatic method to prove it for the vertex contributions 
in Appendix \ref{appencomp}.

 \section{General Vertex Contributions to EE}
\label{s:vertex}
In this section, we extend the analysis of vertex contributions to EE from the $\phi^4$ interaction to more general cases.
\subsection{$\phi^6$ scalar field theory}
\begin{figure}[t]
\centering
\includegraphics[width=9cm]{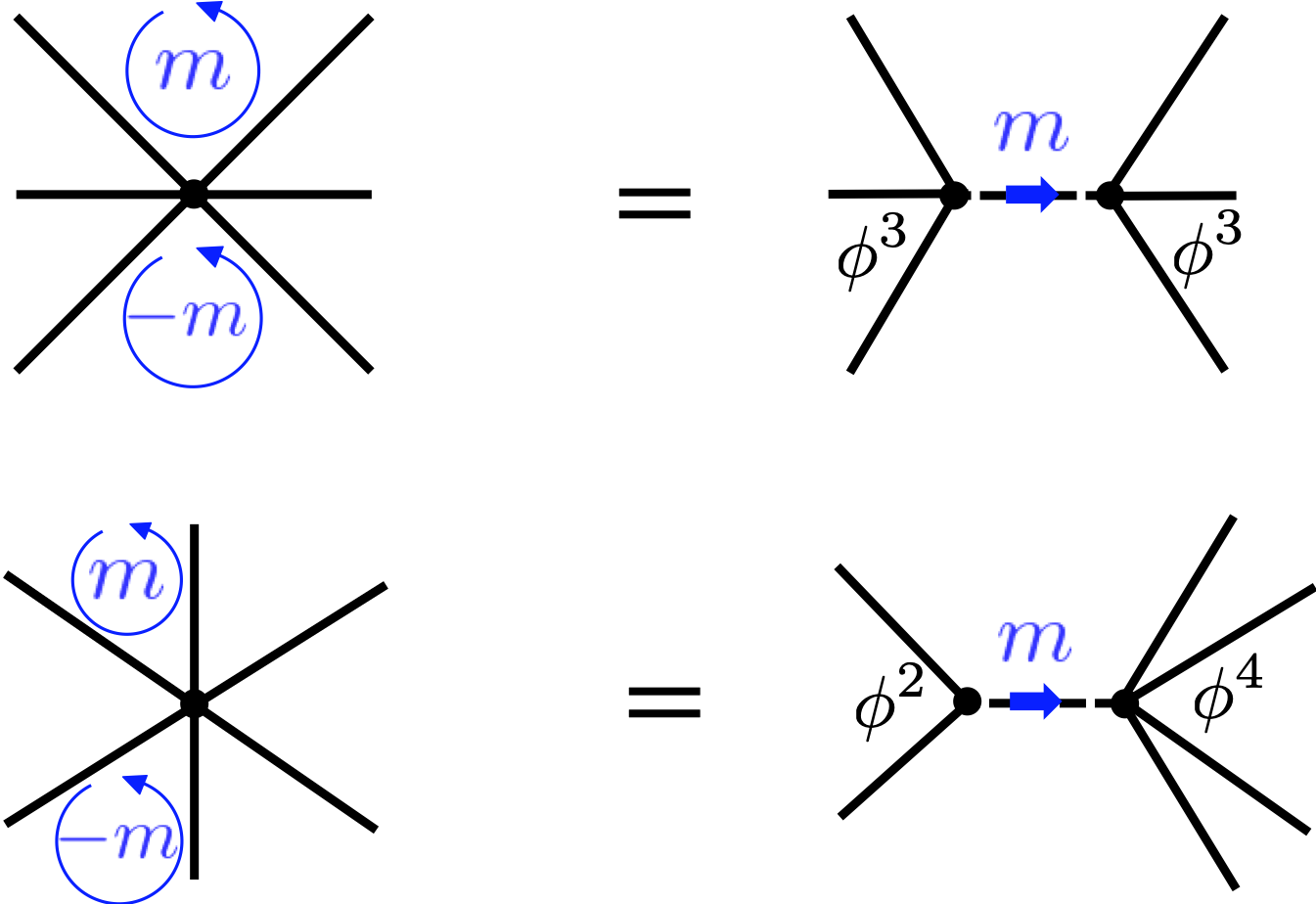}
\caption{
Two different composite operators appear by
opening the $\phi^6$ vertex. Each flux configuration
corresponds to twisting the propagator of the respective composite operator. 
}
\label{f:6point-vertex}
\end{figure}

First, let us consider the $\phi^6$ interaction,
\aln{
{\cal L}_{\mathrm{pot}}=\frac{\lambda_6}{6 } \phi^6.
}
In this case, we  have  two types of vertex configurations\footnote{
{
The $\phi^6$ interaction will induce $\phi^4$ interaction by contracting
two $\phi$'s, but in this section, we simply set it zero by renormalization
and do not consider contributions to EE from such diagrams as the vertex contributions at this stage.
A model containing both of $\phi^4$ and $\phi^6$
interaction vertices are studied in the next section. 
 }
}
 as drawn in Fig.\ref{f:6point-vertex}
and need to introduce three types of composite operators, $\phi^2$, $\phi^4$, and $\phi^3$, 
to extract all the vertex contributions to EE. Since the theory has $\mathbb{Z}_2$ invariance under $\phi \rightarrow -\phi$, 
the $\mathbb{Z}_2$-even operators,  $\phi^2$ and $\phi^4$, are mixed with themselves 
while the $\mathbb{Z}_2$-odd operator $\phi^3$ is mixed with the fundamental field $\phi$.
Therefore,  the propagator 
contribution in Eq.(\ref{e:EE2PIprop}) needs a  modification. 

First, let us consider the modified propagator contributions in the $\phi^6$ theory. 
\begin{figure}[t]
\centering
\includegraphics[width=6cm]{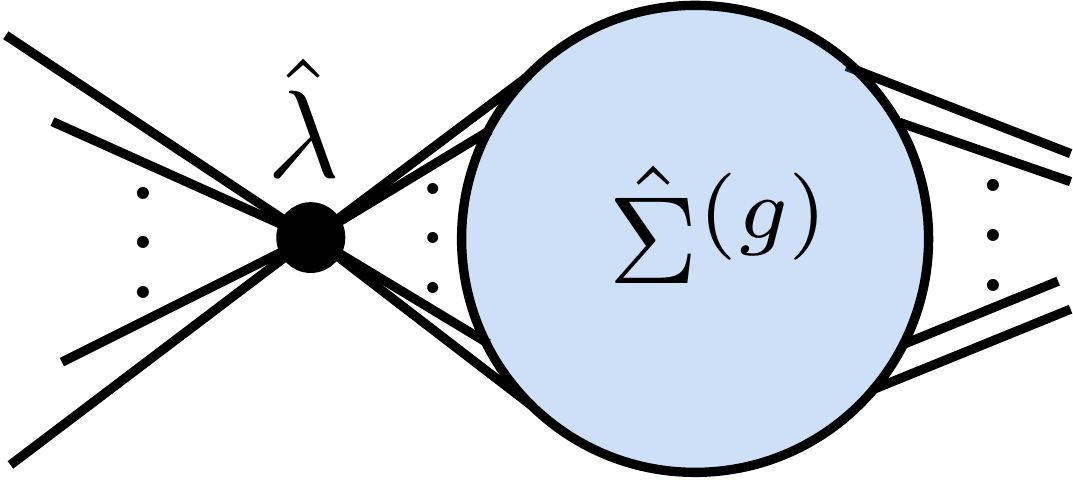}
\caption{
A Schwinger-Dyson type diagram to represent  mixings between different operators. All possible composite operators
are assigned to each dotted part. $\hat\lambda$ is a matrix-valued vertex  
and $\Sigma^{(g)}$ is a generalized 1PI (g-1PI) self-energy with respect to
the composite operators. 
}
\label{f:13mixing}
\end{figure}
Such contributions come from 1-loop type diagrams of
mixed correlations of $\phi$ and $\phi^3$ operators. 
They are given by (Fig.\ref{f:13mixing})
\aln{
 S_{\mathbb{Z}_2\text{-odd}}
 =   &  - \frac{V_{d-1} }{6} 
 \int \frac{d^{d-1}k_{\parallel}} {2(2 \pi)^{d-1}} \tr \log   
 \left[\hat{G}_0^{-1} - \hat\lambda\hat{\Sigma}^{(g)}  \right],
 \label{EEphi6-Z2odd}
 }
 where 
 \aln{
  \hat{G}_0 =&   \left( \begin{array}{cc} G_0  & 0 \\0 & 
  1  \end{array}\right)  , 
  \   \ 
 \hat\lambda= \left(\begin{array}{cc}1  & 0   \\ 0  & -10 \lambda_6  /3 \end{array}\right) ,   
  \   \
  \hat{\Sigma}^{(g)} =    \left(\begin{array}{cc}  \Sigma^{(g)}  \, & \Sigma^{(g)}_{\phi\phi^3}
  \\ \Sigma^{(g)}_{\phi^3\phi} & \Sigma^{(g)}_{\phi^3\phi^3}\, \end{array}\right) .
  \label{GLS}
 }
 It is a natural generalization of Eq.(\ref{GLS1}) including an operator mixing. 
The diagonal component of $\hat{G}_0$ is the bare propagators of $\phi$ and $\phi^3$ operators, respectively.
$\hat\lambda$ is a matrix whose matrix element 
represents the coefficients of opening the $\phi^6$ vertex.
{The coefficient for $\phi^3$ to $\phi^3$ in $\hat\lambda$ is given by $1/6 \times _6 C_3$.} 
$\hat{\Sigma}_{}^{(g)} = \hat{\Sigma}_{}^{(g)} ({\bm k}=\bm{0}, k_\parallel)$ is the $\mathbb{Z}_2$-odd
g-1PI function
\footnote{
{In the $\mathbb{Z}_2\text{-odd}$ set of operators, 
the $[\phi^5]$ operator does not appear in the  mixing, though
the $\phi^6$ vertex can be decomposed into $\phi$ and $\phi^5$. 
It is because a diagram with $\langle\, \phi\, [\phi^5]\, \rangle$ is not 1PI while the g-1PI is 1PI as well in the ordinary sense. } 
}.  
Namely, it consists of 1PI diagrams that do not contain beads diagrams shown
 in Fig.\ref{f:g1PI}.
 Such a generalization of the 1PI concept is mandatory since, in calculating 
the vertex contributions to EE,  
we need to open a vertex to take account of various  channel contributions 
and  special care of the beads diagram in Fig.\ref{f:g1PI} is necessary. 
This is the reason why we have generalized the concept of 1PI.

\vspace{5mm}
The above discussions can be straightforwardly extended to the contributions from 
$\mathbb{Z}_2$-even operators,  $\phi^2$ and $\phi^4$. 
This case is simpler because the bare Green function is unity; $G^{(g)}=1$. 
Then, we have the same matrix form
\aln{
S_{\mathbb{Z}_2\text{-even}} = &  - \frac{V_{d-1} }{6} 
 \int^{1/\epsilon} \frac{d^{d-1}k_{\parallel}} {2(2 \pi)^{d-1}} \tr \log (\hat{1}-
 \hat\lambda 
 \hat{\Sigma}^{(g)} ),
 \label{EEphi6-Z2even}
}
where,  in this case,  matrices are given by
\aln{
 \hat\lambda
  = &
 \left(\begin{array}{cc}0 & -5 \lambda_6/2  \\-5 \lambda_6/{2} & 0\end{array}\right), \ \ 
 \hat{\Sigma}^{(g)} =
\left(\begin{array}{cc}\Sigma^{(g)}_{\phi^2\phi^2}&  \Sigma^{(g)}_{\phi^2\phi^4}
\\ \Sigma^{(g)}_{\phi^4\phi^2}& \Sigma^{(g)}_{\phi^4\phi^4}  \end{array}\right).
\label{X}
}
The coefficient comes from $5/2 =1/6 \times {}_6 C_2$. 
It is a $2 \times 2$ matrix generalization of Eq.(\ref{e:EEcomp}).
The g-1PI self-energy $\hat\Sigma^{(g)}$ does not contain 
beads diagrams, especially diagrams connected by 
the $\phi^6$ vertex decomposed into $\phi^2$ and $\phi^4$. 

Note that  EE of Eqs.(\ref{EEphi6-Z2odd}) and (\ref{EEphi6-Z2even}) written in terms of the g-1PI functions 
can be rewritten  in terms of the renormalized correlation functions
as in the $\phi^4$ case of  Eqs.(\ref{e:EEprop}) and (\ref{G=sumofg1PI}).
The only difference is that we now have operator mixings and the relationship becomes more complicated. 
Let us explicitly check it for the $\mathbb{Z}_2$-odd case of Eq.(\ref{EEphi6-Z2odd}).
It is rewritten as
\aln{
S_{\mathbb{Z}_2\text{-odd}}&=\frac{V_{d-1}}{6}\int^{1/\varepsilon}\frac{d^{d-1}k_\parallel}{2(2\pi)^{d-1}}\mathrm{tr}\,\mathrm{ln}\left(\hat{G}_0\,\frac{1}{\hat{1}-\hat{\lambda}_{}\hat{\Sigma}_{}^{(g)}\hat{G}_0}\right)\nonumber\\
&=\frac{V_{d-1}}{6}\int^{1/\varepsilon}\frac{d^{d-1}k_\parallel}{2(2\pi)^{d-1}}\mathrm{tr}\,\mathrm{ln}\left(\hat{G}_0+\hat{G}_0\hat{\lambda}_{}\hat{\Sigma}_{}^{(g)}\hat{G}_0+\hat{G}_0\hat{\lambda}_{}\hat{\Sigma}_{}^{(g)}\hat{G}_0\,\hat{\lambda}_{}\hat{\Sigma}_{}^{(g)}\hat{G}_0+\cdots\right).
}
Writing the inside of the parenthesis as $\tilde{G}$, its matrix elements are given by
\aln{
(\tilde{G})_{11}&=G_0+G_0\Sigma^{(g)}G_0+G_0\Sigma^{(g)}_{\phi\phi^3}\left(-\frac{10}{3}\lambda_6\right)\Sigma^{(g)}_{\phi^3\phi}G_0+G_0\Sigma^{(g)}G_0\Sigma^{(g)}G_0+\cdots,\\
(\tilde{G})_{12}&=G_0\Sigma^{(g)}_{\phi\phi^3}+G_0\Sigma^{(g)}G_0\Sigma^{(g)}_{\phi\phi^3}+G_0\Sigma^{(g)}_{\phi\phi^3}\left(-\frac{10}{3}\lambda_6\right)\Sigma^{(g)}_{\phi^3\phi^3}+\cdots,\\
(\tilde{G})_{21}&=\left(-\frac{10}{3}\lambda_6\right)\left(\Sigma^{(g)}_{\phi^3\phi}G_0+\Sigma^{(g)}_{\phi^3\phi}G_0\Sigma^{(g)}G_0+\Sigma^{(g)}_{\phi^3\phi^3}\left(-\frac{10}{3}\lambda_6\right)\Sigma^{(g)}_{\phi^3\phi}G_0+\cdots,\right),\\
(\tilde{G})_{22}&=1+\left(-\frac{10}{3}\lambda_6\right)\left(\Sigma^{(g)}_{\phi^3\phi^3}+\Sigma^{(g)}_{\phi^3\phi}G_0\Sigma^{(g)}_{\phi\phi^3}+\Sigma^{(g)}_{\phi^3\phi^3}\left(-\frac{10}{3}\lambda_6\right)\Sigma^{(g)}_{\phi^3\phi^3}+\cdots,\right).
}
We can explicitly see that the sum of g-1PI's in each matrix element is combined into 
 the ordinary 1PI functions {$\Sigma$'s}, and hence can be written by the 
 renormalized correlation functions as
\aln{
(\tilde{G})_{11}&=G_0+G_0\Sigma G_0+G_0\Sigma G_0\Sigma G_0+\cdots    
=G,\\
(\tilde{G})_{12}&=(G_0+G_0\Sigma G_0+G_0\Sigma G_0\Sigma G_0+\cdots)\Sigma_{\phi\phi^3} 
=G_{\phi\phi^3},\\
(\tilde{G})_{21}&=\left(-\frac{10}{3}\lambda_6\right)\Sigma_{\phi^3\phi}(G_0+G_0\Sigma G_0+G_0\Sigma G_0\Sigma G_0+\cdots) 
=-\frac{10}{3}\lambda_6G_{\phi^3\phi},\\
(\tilde{G})_{22}&=1+\left(-\frac{10}{3}\lambda_6\right)\left(\Sigma_{\phi^3\phi^3}+\Sigma_{\phi^3\phi}G\Sigma_{\phi\phi^3}\right)
=1-\frac{10}{3}\lambda_6G_{\phi^3\phi^3}.
}
As a result,  Eq.(\ref{EEphi6-Z2odd}) can be summarized as
\aln{
 S_{\mathbb{Z}_2\text{-odd}}
 =   &  \frac{V_{d-1} }{6} 
 \int^{1/\epsilon} \frac{d^{d-1}k_{\parallel}} {2(2 \pi)^{d-1}} \tr \log   
 \left[ \tilde{I}+ \hat\lambda_{}\hat{G}\right],
 \label{eq:gen-to-ren}
}
where
\aln{
\tilde{I}=\left(\begin{array}{cc}
0 & 0 \\
0 & 1
\end{array}\right)
, 
\   \ 
{\hat\lambda= \left(\begin{array}{cc}1  & 0   \\ 0  & -10 \lambda_6  /3 \end{array}\right) ,   }
\   \
\hat{G}=\left(\begin{array}{cc}
G                   & G_{\phi\phi^3} \\
G_{\phi^3\phi}  & G_{\phi^3\phi^3}
\end{array}\right).
}
The same discussion can be applied to  Eq.(\ref{EEphi6-Z2even}).
This gives  an alternative, unified formula for EE in terms of the renormalized Green functions. 
\subsection{$\phi^4 +\phi^6$  theory and further generalizations}
Let us generalize a bit more and consider a case when the Lagrangian contains two interaction terms
\aln{
{\cal L}_{pot}=\frac{\lambda_4}{4 } \phi^4 + \frac{\lambda_6}{6 } \phi^6 .
}
As in the $\phi^6$ theory, we need to consider three composite operators, $\phi^2$, $\phi^4$, and $\phi^3$,
in order to take into account contributions to EE from these vertices. 
Again, we have $\mathbb{Z}_2$ invariance and EE is a sum of $\mathbb{Z}_2$-even and odd contributions. 
The $\mathbb{Z}_2$-odd contribution is given by
\aln{
 S_{\mathbb{Z}_2\text{-odd}}
 =  & - \frac{V_{d-1} }{6} 
 \int^{1/\epsilon} \frac{d^{d-1}k_{\parallel}} {2(2 \pi)^{d-1}} \tr \log   
 \left[ 
 \hat{G}_0^{-1} -
\left(\begin{array}{cc}1  & 0
\\ 0 
  & -10 \lambda_6/3 \end{array}\right)
 \left(\begin{array}{cc}  \Sigma^{(g)}_{}  & \Sigma^{(g)}_{\phi\phi^3}
  \\  \Sigma^{(g)}_{\phi^3\phi}  & \Sigma^{(g)}_{\phi^3\phi^3} \end{array}\right) 
 \right],
  \label{EEphi46-Z2odd}
 }
 where $\hat{G}_0$ is the same as in Eq.(\ref{GLS})
 while $\mathbb{Z}_2$-even contribution is given by 
 \aln{ 
S_{ \mathbb{Z}_2\text{-even}} = & - \frac{V_{d-1} }{6} 
 \int^{1/\epsilon} \frac{d^{d-1}k_{\parallel}} {2(2 \pi)^{d-1}} 
 \tr \log \left[ \hat{1}-
 \left(\begin{array}{cc} -3\lambda_4/2 & -5 \lambda_6/2  \\-5 \lambda_6/2 & 0\end{array}\right)
\left(\begin{array}{cc} \Sigma^{(g)}_{\phi^2\phi^2}  &  \Sigma^{(g)}_{\phi^2\phi^4}
\\ \Sigma^{(g)}_{\phi^4\phi^2}  & \Sigma^{(g)}_{\phi^4\phi^4} \end{array}\right)  
 \right]  .
 \label{EEphi46-Z2even}
}

\vspace{5mm}
Now a generalization to e.g.  $\phi^{2n}$ vertices with higher $n$ is evident.
The propagator and vertex contributions to EE are unified to be written in a  matrix form as 
Eq.(\ref{EEphi6-Z2odd}):
\aln{
 S_{EE}
 =   & - \frac{V_{d-1} }{6} 
 \int^{1/\epsilon} \frac{d^{d-1}k_{\parallel}} {2(2 \pi)^{d-1}} \tr \log   
 \left[ (\hat{G}_0^{-1} -  \hat\lambda \hat{\Sigma}^{(g)}  ) ({\bm k}=0, k_\parallel) \right].
 \label{EE-generalform2}
 }
 The size of matrices becomes larger as a larger number of operators are mixed and 
each set of mixed operators forms a block diagonal component. 
$\hat{G}_0$ is a diagonal matrix whose entry is mostly 1 except the fundamental field. 
$\hat{\lambda}$ represents a mixing among operators via vertices while $\hat{\Sigma}^{(g)}$ 
represents amputated correlators of all the fundamental and composite operators. 
The notion of the g-1PI is also extended to exclude all the beads diagrams 
constructed by all the vertices along with the ordinary non-1PI diagrams. 
This form of EE contains all the contributions from the propagators and the vertices. {We provide the derivation in Appendix \ref{appencomp}.}

An essential point is that we can rewrite Eq.(\ref{EE-generalform2}) 
in terms of the renormalized correlation functions in the same manner as in {Eq.\eqref{eq:gen-to-ren}} as 
\aln{
S_{EE}=\frac{V_{d-1}}{6}\int^{1/\varepsilon}\frac{d^{d-1}k_\parallel}{2(2\pi)^{d-1}}\mathrm{tr}\,\mathrm{ln}\left(\tilde{I}+{\hat{\lambda}} 
 \hat{G}\right).
\label{EEren}
}
Here, $\tilde{I}=\text{diag}(0,1,\cdots,1)$, $\hat{G}$ is the matrix form of the correlators of operators, and we have arranged the elements of the matrices so that the first line and first column involve the fundamental field $\phi$. 
The size of the matrices is finite as far as there is a finite number of vertices. 
In the $\phi^n$-theory,
 we need to consider only the composite operators $[\phi^j]$ with $j\leq n-2$, which appear to open vertices. 

\subsection{Derivative interactions}
Special care is necessary for generalizations with derivative interactions 
since composite operators with Lorentz indices appear. 
Let us consider the following interaction as an example, 
\aln{
{\cal L}_{pot}= \frac{\lambda_{{\partial}}}{4}(\phi \partial \phi)^2 . 
}
In this case,  the two types of scalar composite operators,  $[\phi^2]$ and $[(\partial \phi)^2]$,  as well as a
spin-1 operator $[\phi \partial_\mu \phi]$ appear from an opened vertex. 
Since the spin-1 operator does not mix with either $\phi$ or $[\phi^2]$ 
or $[(\partial \phi)^2]$, we can separately study its contribution to EE.  
Thus we have three block-diagonal sectors.

The spin-0 sectors can be treated as  before. 
Thus let us focus on the spin-1 sector. The formula Eq.(\ref{EE-generalform2}) gets a bit modified since EE of 
 a spinning field is different from that of a scalar field due to the rotation
of the internal spin induced by $\mathbb{Z}_M$ twist 
 and hence an extra phase appears in evaluating EE \cite{He:2014gva}. 
 The operator $J_\mu:= [\phi \partial_\mu \phi]$ is decomposed into its two-dimensional part ${\bm J}$
and $(d-1)$-dimensional part $J_i$. The latter is a scalar on 
{the two-dimensional spacetime normal to the boundary} and can be treated 
as in Eq.(\ref{EE-generalform2}). 
On the other hand, the contribution to EE from the 2-dimensional vector ${\bm J}$ is modified. 
From  Eq.(2.21) in \cite{He:2014gva}, the coefficient of EE is proportional to 
\aln{
c_{\rm eff}^{\rm boson}(s)=\frac{1}{4} \frac{\partial J(s, M)}{\partial M} \Big|_{M=1} =\frac{1}{6} -\frac{|s|}{2}
}
for a bosonic field with spin {$s$}
. This coefficient $c_{\rm eff}$ replaces the coefficient of
$1/6$ in front of Eq.(\ref{e:EE2PIprop}). 
Thus for $(d+1)$-dimensional vector $J_\mu$, the total coefficient is 
given by {$(d-1)/6+2(1/6-1/2)=(d-5)/6.$ }
Therefore the propagator and vertex contributions to 
EE with this derivative interaction is given by either of the following two forms, 
\aln{
 S_{EE}
 =   & - 
 {\frac{V_{d-1}}{6}}
 \int^{1/\epsilon} \frac{d^{d-1}k_{\parallel}} {2(2 \pi)^{d-1}} \tr \left( S \log   
 \left[ \hat{G}_0^{-1}-  \hat{\lambda} \hat{\Sigma}^{(g)}  \right] \right)\nonumber\\
 = & 
 {\frac{V_{d-1}}{6}}
 \int^{1/\epsilon} \frac{d^{d-1}k_{\parallel}} {2(2 \pi)^{d-1}} \tr \left( S \log   
 \left[ \tilde{I}+  \hat{\lambda} \hat{G}  \right] \right),
 \label{EE-derivative} 
 }
 where 
 \aln{
S&= 
\left(\begin{array}{cccc}1 & 0 & 0 & 0 \\0 & 1 & 0 & 0 \\0 & 0 & 1 & 0 \\0 & 0 & 0 & (d-5) \end{array}\right), \ 
\tilde{I}= \left(\begin{array}{cccc} 0 & 0 & 0 & 0 \\0 & 1 & 0 & 0 \\0 & 0 & 1 & 0 \\0 & 0 & 0 & 1\end{array}\right),
\ \ 
\hat{\lambda } = \left(\begin{array}{cccc}1 & 0 & 0 & 0 \\0 & 0 & -\lambda_{{\partial}} & 0 \\0 & -\lambda_{{\partial}} & 0 & 0 
\\0 & 0 & 0 & -\lambda_{{\partial}}/2\end{array}\right), \nn 
 \hat{G}^{}  &= \left(\begin{array}{cccc}
 {G} & 0 & 0 & 0 \\
 0 & G_{\phi^2\phi^2}  & G_{\phi^2(\partial\phi)^2} & 0 \\
 0 &  G_{(\partial\phi)^2\phi^2} & G_{(\partial\phi)^2(\partial\phi)^2} & 0 \\
 0 & 0 & 0 & G_{(\phi\partial_\mu\phi)(\phi\partial^\mu\phi)}\end{array}\right) .
}
$S$ is an additional coefficient due to the spin. Here we have summed over $(d+1)$-dimensional vector contributions,
but generally speaking,  it is more convenient to write a matrix corresponding to 
each irreducible representation of the 2-dimensional rotation with spin $s$. 
According to \cite{He:2014gva}, the coefficient $c$ for fermions with odd half-integer spin $s$ is given by
$c_{\rm eff}^{\rm fermion}(s) = -1/3$. Thus if we treat each 2-dimensional spin component as an independent
field, the diagonal component of the matrix $S{/6}$ is given by $c_{\rm eff}^{\text{boson/fermion}}(s)$
for each spin $s$ field. 

\section{IR behavior of EE and Wilsonian effective action}
\label{Wilson}
So far we have succeeded to extract  contributions to EE, particularly 
from propagators of fundamental fields  and  vertices, and 
its most general form is given by the unified matrix form in Eq.(\ref{EE-derivative}).  
Thus if we can calculate  correlators $\hat\Sigma^{(g)}$ or $\hat{G}$, we can obtain
their contributions to EE. 
These contributions to EE are, however, a part of the whole EE in the framework of 
the $\mathbb{Z}_M$ gauge theory on Feynman diagrams 
and we wonder how we can extract
the other contributions to EE. In this section, we give a conjecture that 
the IR part of EE is 
exhausted by summing all
the vertex contributions (together with propagator contributions)
constructed from the IR Wilsonian effective action. 
\subsection{More properties of vertex {contributions} to EE}
First, 
note that  the leading order term of the vertex contribution corresponding to 
composite operators $\{{\cal O}_n\}$ 
 is perturbatively given by expanding the logarithm as
\aln{
S_{\rm vertex} &= \frac{V_{d-1}}{12} \int  \frac{d^{d-1}k_{\parallel}} {(2 \pi)^{d-1}} \tr \hat\lambda   \hat\Sigma^{(g)} ({\bm k}=0, k_\parallel)
\nn
&= \frac{V_{d-1}}{12} \int  d^{2}{\bm r} \sum_{m,n} (\hat\lambda)_{mn}   \langle{\cal O}_n(-\frac{{\bm r}}{2}, x_\parallel=0) 
{\cal O}_m (\frac{{\bm r}}{2}, x_\parallel=0) \rangle^{(g)} .
}
The integral in the second line reflects the property of a twisted propagator
that its center coordinate is pinned at the boundary $x_\parallel=0$ with two loose ends. 
For more detailed discussions, see  \cite{Iso:2021vrk, Iso:2021rop}. 
{For instance, when ${\cal O}=[\phi^2]$,} the leading perturbative term 
is given by using the renormalized  propagator $G$ of the fundamental field $\phi$ as 
\aln{
S_{\rm vertex} \sim 
\frac{V_{d-1}}{6} \int  d^{2} {\bm r} \left(-\frac{3\lambda_4}{2}\right)  \ G({\bm r},0)^2 .
\label{EE-G2}
}
$-3\lambda_4/2$ is the component of $\hat{\lambda}$ which associates $[\phi^2]$ to $[\phi^2]$.
If we consider an operator such as ${\cal O}=[\phi^n]$, 
the integrand is proportional to $G({\bm r},0)^n$ and decays faster for a larger $n$.  
This means that at least perturbatively, 
higher-dimensional composite operators tend to 
contribute less to EE. 

Another important point to note in Eq.(\ref{EEren}),  particularly for its vertex part, is that
if some composite operators in $\hat{\lambda} \hat{G}$ dominates $1$ in the logarithm in a strong coupling region, 
the contribution from the composite operator can be approximated as
\aln{
\mathrm{tr}\,\log (\hat{1}+ \hat\lambda \hat{G}) \sim   \mathrm{tr}\,\log (\hat{G})
} 
up to a constant depending on the coupling constant. 
Then, EE can be written as a logarithm of renormalized correlators similar to the fundamental field. 
There is no explicit dependence on the coupling constant {other than the overall factor} and
its dependence is only given through the renormalization of correlators. 

\subsection{Wilsonian RG and EE: free field theories}
Now we discuss the issue of other contributions to EE besides the propagators and vertices. 
For this purpose, it is convenient to utilize the concept of the 
 Wilsonian renormalization group (RG)  to the effective field theory in the IR region \cite{Wilson:1973jj,Polchinski:1983gv}.\footnote{A modern approach for the Wilsonian RG is given by the functional RG method \cite{Wetterich:1992yh, Morris:1993qb}.}
In the Wilsonian RG, we first divide the momentum domain into low and high regimes.
Schematically, 
\aln{
{
	k} \in [0, \Lambda] = [0, e^{-t}\Lambda] + [e^{-t}\Lambda, \Lambda]
}
with $t>0$ and then, integrate quantum fluctuations over the high regimes. 
Then, we rescale the momentum ${
	k} \rightarrow {
	k^\prime} = e^{t}{
	k}$  so that 
${
	k^\prime} \in [0, \Lambda]$. In this procedure, the original parameters in the action
are renormalized, e.g., 
\aln{
m \rightarrow m^\prime, \hspace{5mm} \lambda_4 \rightarrow \lambda_4^\prime. 
}
In addition, new interaction terms appear, e.g., in {the $\phi^4$ theory in} $(3+1)$ dimensions, 
\aln{
\lambda_6  \frac{\phi^6}{\Lambda^2},  \hspace{5mm}  \lambda_{{\partial}} \frac{(\phi \partial \phi)^2}{\Lambda^2}, 
\hspace{5mm}  \lambda_8  \frac{\phi^8}{\Lambda^4}  \cdots .
}

\vspace{5mm}
First, let us look at what happens for a free theory. For a free scalar field with a mass $m$, EE is simply given by 
\aln{
S_{\text{EE}}(\Lambda) & =- \frac{V_{d-1} }{12} 
 \int^{\Lambda} \frac{d^{d-1}k_{\parallel}} {(2 \pi)^{d-1}}  
  \log \left[ (k_\parallel^2 +m^2) / \Lambda^2 \right] .
  \label{e:EEprop-free}    
 }
 By integrating the high momentum region, nothing happens except fluctuations of that region
 are discarded: 
 \aln{
 S_{\text{EE}} (e^{-t}\Lambda) & =- \frac{V_{d-1} }{12} 
 \int^{e^{-t}\Lambda} \frac{d^{d-1}k_{\parallel}} {(2 \pi)^{d-1}}  
  \log \left[ e^{2t}(k_\parallel^2 +m^2) / \Lambda^2 \right] .
 }
 Then, we rescale the momentum as $k^\prime = e^t k$ to obtain 
 \aln{
  S_{\text{EE}}^\prime (\Lambda)  =- 
 \frac{V_{d-1} }{12} 
 \int^{\Lambda} \frac{d^{d-1}k^\prime_{\parallel}}{e^{(d-1)t} \times (2 \pi)^{d-1}}  
  \log \left[ (k^{\prime 2}_\parallel + e^{2t} m^{ 2}) / \Lambda^2 \right] .
  \label{EE-IR-free}
 }
Of course, for a free field theory, it is equal to Eq.(\ref{e:EEprop-free}) with the integration range $[0, e^{-t}\Lambda]$. 
For an interacting theory, it is different since high and low momentum {modes} are entangled. 
We continue the integration over high momentum modes until $e^{-t}\Lambda = m$. 
Then, EE is given by Eq.(\ref{e:EEprop-free}) with the integration range $[0,  m]$.
It gives the  IR part of the EE at the scale $m$, and
the discarded parts in higher momentum are  UV cut-off dependent.  
By performing the momentum integration, 
{the EE at the scale $m$} is now given by 
\aln{
  S_{\text{EE}}^{\rm IR} (m) 
  	& {\equiv - \frac{V_{d-1} }{12} 
  	\int^{m} \frac{d^{d-1}k_{\parallel}} {(2 \pi)^{d-1}}  
  	\log \left[ (k_\parallel^2 +m^2) / \Lambda^2 \right] } \\
  & =
 \frac{N_{\rm eff} V_{d-1} }{12}
  \,  m^{d-1}\log \left[  \tilde\Lambda^2/ m^2 \right] ,
  \label{EE-IR-free2}
 }
where $\tilde{\Lambda}$ is proportional to the  UV cutoff 
as $\tilde{\Lambda}= \Lambda \exp[\Phi(-1,1,\frac{d+1}{2})/2] /\sqrt{2}$. $\Phi(z,s,\alpha)\equiv \sum_{n=0}^\infty z^n/(n+\alpha)^s$ is the Lerch transcendent.
For example, in $d=3$, it is given  by $\tilde{\Lambda}={e^{1/2} \Lambda}/{2}$. 
{Eq.\eqref{EE-IR-free2} coincides with the ordinary universal term in even spacetime dimensions.} 
 $V_{d-1}$ is the area of the boundary and 
\aln{
 N_{\rm eff}=  \left(\frac{1}{2}\right)^{d-1} \frac{1}{ \pi^{(d-1)/2} \Gamma((d+1)/2)} 
} 
is the effective {number of} degrees of freedom that can contribute to  EE {in the IR}. 
The result of Eq.(\ref{EE-IR-free2}) 
{indicates} that the universal part of EE originates in the quantum correlations
of fields whose length scale is larger than the typical correlation length $\xi=1/m$ of the system.
$S_{\text{EE}}^{\rm IR} (m)$ becomes larger for smaller masses $m$. 
\subsection{Wilsonian RG and EE:  interacting field theories}
In the free case, the Wilsonian RG  can extract the {IR} behavior of EE that is independent of the UV
cutoff. 
In the Wilsonian RG,  quantization is gradually performed from high momentum to low, and
 in the IR limit, all fluctuations are integrated out so that
all the loop effects are incorporated in  the Wilsonian effective action (EA).
{The Wilsonian EA} becomes more and more complicated as radiative corrections are gradually taken into account. 
Thus we can expect that all the contributions to EE are encoded in the Wilsonian EA.
We conjecture that EE is given by a sum of all the {propagator and} vertex contributions in the Wilsonian EA.  

\begin{figure}[t]
\centering
\hspace*{-1.2cm}
\includegraphics[width=18.7cm]{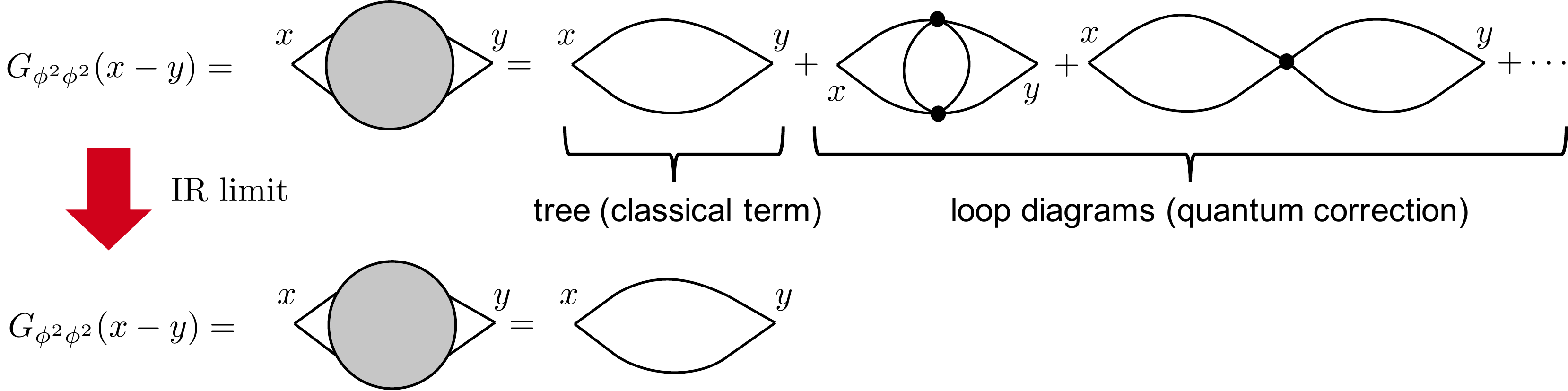}
\caption{
The upper figure shows Feynman diagrams constituting $G_{\phi^2\phi^2}$.
Only the leading diagram survives in the IR limit. 
}
\label{f:vertexWilson}
\end{figure}
In the following, we focus on the IR limit of the Wilsonian EA. 
Let us recall a simple case in Eq.(\ref{e:EEcomp2}). 
The correlator $G_{\phi^2\phi^2}$ is graphically given by the upper figure of 
Fig.\ref{f:vertexWilson} and
the first term is  given by Eq.(\ref{EE-G2}). 
This diagram is present even in the IR limit where all the fluctuations are integrated out since
it is simply connected by the propagators of the fundamental field. 
The other terms vanish in the IR limit of the Wilsonian RG
since they are quantum corrections to the first classical term. 
After all the fluctuations are integrated out, further quantum corrections should be absent because
such effects are already absorbed in the Wilsonian EA. 
Thus we expect that the vertex contributions of the composite operator, e.g. $\phi^2$, are drastically simplified in the IR limit in which we can 
replace the Green function $G_{\phi^2\phi^2}$ by the leading diagrams as shown in the lower figure of Fig.\ref{f:vertexWilson}.
After all, the vertex contribution in Eq.(\ref{e:EEcomp2}) becomes 
\aln{
S_{\text{vertex}}= 
		 - \frac{V_{d-1}}{12}
		\int^{\Lambda}\frac{d^{d-1}k_\parallel}{(2\pi)^{d-1}}\mathrm{log}\,\left[1 + 
		3 \lambda_4  
	       \int\frac{d^{d+1}p}{(2\pi)^{d+1}}G(p)G(-\bm{p},k_\parallel-p_\parallel)   
		\right] 
		\label{e:EEvertexRG}
}
in the IR limit. The coupling constant $\lambda_4$ is the renormalized one since it is a coefficient of 
Wilsonian EA in the IR limit.\footnote{
We have already taken quantum fluctuations into account and eliminated the UV divergences in coupling constants and observables in the IR limit, but another UV divergences appear in the calculation of EE since 
we need to sum all the momentum modes. It is also necessary even for the free theory 
and indeed we extracted the IR universal part by subtracting cutoff dependent terms. }
As in the free case, we separate the vertex contributions into  
{IR and UV parts}.
The IR part is defined similarly by restricting the integration range from $k_\parallel \in [0, \Lambda]$
to $[0, m]$.  
Instead, we may integrate  up to $1/\xi_{\phi^2}$
where  $\xi_{\phi^2}$ is the correlation length of the operator $[\phi^2]$. 
The difference is a matter of definition of the IR universal part of EE and we need a 
precise prescription to subtract the cutoff dependent terms in EE. 
For example, we may  take a variation with respect to the mass $m$ and then integrate to obtain the universal 
part of EE. In this definition, we need to know how $\xi_{\phi^2}$ and $m$ are related. 
{This is under investigation in moment.}

In general, of course, we need to take operator mixings into account
but the generalization is straightforward. 
We will investigate 
more detailed behaviors of EE in the infrared limit of the Wilsonian EA for a concrete model.
The final question is whether there are contributions to EE other than the  vertex contributions
in the Wilsonian EA. 
In the formulation of EE based on the $\mathbb{Z}_M$ gauge theory on Feynman diagrams, 
vertex contributions are only a part of all the contributions to EE. 
But, in the IR limit of Wilsonian RG, all the quantum fluctuations are integrated out 
and  we do not need to evaluate loop diagrams: all the Feynman diagrams are tree diagrams. 
Thus  the vertex contributions, as well as the propagator contributions, to EE must
suffice {for} the IR behavior of EE. 
 We will investigate further issues of the RG flow of EE in a separate paper~\cite{IMS4}.


 \section{Conclusions}
\label{s:discussion}
This is the third paper in a series of our investigations on EE in interacting field theories
based on the notion of the $\mathbb{Z}_M$ gauge theory on Feynman diagrams, 
proposed in~\cite{Iso:2021vrk} and extended in~\cite{Iso:2021rop}. 
In the previous papers, we have focused on two important contributions to EE; 
one from the propagators of the fundamental field and another from vertices
which can be interpreted as correlations of composite operators.
In this paper, we have further extended the results to include effects of mixings of  various composite operators
as well as the original fundamental fields.
The final formula of EE is  given in a unified matrix form.
We then discuss an implication to the IR behavior of EE
 from the Wilsonian RG approach to effective field theories. 
We conjecture that the IR  part of EE in interacting field theories is given by a sum of 
all vertex contributions in the Wilsonian effective action. 
In this context, it is interesting to look at the relation to the variational method of EE \cite{Cotler:2015zda, Fernandez-Melgarejo:2020utg, Fernandez-Melgarejo:2021ymz, Fernandez-Melgarejo:2020fzw}. 
In this approach, EE of interacting field theories is expressed 
in terms of a   non-Gaussian deformation 
of the Gaussian vacuum wave function.  This deformation must be related to the vertex contributions
we have found. 


\appendix
\section{Area laws for R\'{e}nyi entropy and entanglement capacity}
\label{appenarea}
In this appendix, we show that the area law\footnote{
	In general, the area law means $O(|\partial A|)$; the quantity scales at most as the area~\cite{Eisert:2008ur}.
} $S_{EE}\propto \mathrm{vol}(\partial A)$ ($\sim V_{d-1}$ in our setup) holds at the level of R\'{e}nyi entropy $S_n\equiv\frac{1}{1-n}\log \Tr \rho_A^n$. We can apply the same discussion performed in our previous papers~\cite{Iso:2021vrk,Iso:2021rop} deriving the area law for EE.
First, to apply the orbifold method introduced in Sec. \ref{s:ZM}, we rewrite $S_{n=1/M}$ in terms of free energy $F^{(M)}$ on $\mathbb{R}^2/\mathbb{Z}_M\times \mathbb{R}^{d-1}$:
\begin{equation}
	S_{1/M}=\frac{1}{M-1}\left(F^{(1)}-MF^{(M)}\right).
	\label{eq:renyi}
\end{equation}
Based on the $\mathbb{Z}_M$ gauge theory on Feynman diagrams introduced in Sec. \ref{s:ZM}, the number of independent twists is given by the number of loops $L$ in the Feynman diagrams even though every propagator is originally twisted. Let us denote the number of initial twists or equivalently the number of propagators by $P$. Twists other than the independent ones can be eliminated by the redundancy at vertices. The number of such redundant twists is given by $P-L=V-1$, where the number of vertices is denoted by $V$. As a result, the $1/M$ factor from each vertex is almost canceled except one by the trivial summation for $(V-1)$ redundant twists. Furthermore, the overall momentum conservation yields the sum of $L$ twisted momenta equals to the original one. In short, any Feynman diagrams contributing to $F^{(M)}$ are expressed as
\begin{align}
		\frac{V_{d-1}}{M}\sum_{\{m\}}\int\prod_{l=1}^L\left[\frac{d^2\bm{p}_l}{(2\pi)^2}\right]I(\{\bm{p}\};\{m\})
		\delta^2 \left(\sum_{l=1}^L (1-\hat{g}^{m_l}) \bm{p}_l \right),
		\label{eq:area-law}
\end{align}
where  
$\sum_{\{m\}}$ is a summation over all twists; each from $0$ to $M-1$. $I(\{\bm{p}\};\{m\})$ is some function of momenta and twists.

When all $m$'s are zero, no momenta are twisted. Such diagrams constitute nothing but $F^{(1)}/M$. Although this contribution in $F^{(M)}$ is proportional to $V_{d+1}$ and seemingly violates the area law, it is canceled in $S_{1/M}$ in 
Eq.\eqref{eq:renyi}. Other configurations of twists include at least one nonzero twist. As a result, the argument of the delta function in Eq.\eqref{eq:area-law} is always nonzero and it combined with $I$ carries a nontrivial dependence in $M$ after the summation over twists. Unless an explicit calculation is done, we do not know the precise $M$ dependence of Eq.(\ref{eq:area-law})
 or $S_{1/M}$. Nevertheless, since terms contributing to $S_{1/M}$ always have nonzero arguments of the delta function, there is no more volume factor other than $V_{d-1}$. 
 If $M$ can be analytically continued to $M=1/n$, this completes the proof of the area law for R\'{e}nyi entropy $S_{n}$. 

The proof above only depends on the technique of Feynman diagrams and thus the area law for R\'{e}nyi entropy is proven for any locally interacting QFTs, given a half space as a subregion.

It is worthwhile to note that the area law for R\'{e}nyi entropy immediately implies the area law for the capacity of entanglement~\cite{PhysRevLett.105.080501,deBoer:2018mzv},
\begin{align}
	C_A&\equiv \lim_{n\rightarrow 1} n^2 \pdv[2]{n}\log \Tr \rho_A^n\nonumber\\
	&=\left.\pdv[2]{n}\left[(1-n)S_n\right]\right\vert_{n\rightarrow 1}\nonumber\\
	&={2}\left.\pdv{S_{1/M}}{M}\right\vert_{M\rightarrow 1}
\end{align}
as well as EE since $C_A$ is linear in R\'{e}nyi entropy. Since $C_A$ is alternatively written as the fluctuation of the modular Hamiltonian $-\log\rho_A$, it is more sensitive to the change of dominant contributions in the replicated geometry and recently discussed in the context of the black hole evaporation~\cite{Kawabata:2021hac,Okuyama:2021ylc,Kawabata:2021vyo}. It is interesting if we can compute such quantities in interacting theories and follow the behavior of higher orders in $M$.

Although the area law itself is intuitive for physicists as entanglement across the boundary $\partial A$ should be dominant for any local QFTs, 
{the proof of this} is difficult; a general proof is known only for gapped systems in $(1+1)$ dimensions~\cite{Hastings:2007iok}. It is remarkable that we can show the area law of both EE and R\'{e}nyi entropy in any locally interacting theories.

As a further generalization, it is intriguing to relax several assumptions and see how the EE and R\'{e}nyi entropy deviates from the area law. In our setup, $\partial A$ is smooth, the interactions are local, and the system is translationally invariant.  
{Some cases are known where the above features are not satisfied and the area law is violated.} For example, when the entangling surface $\partial A$ has a singular geometry, a logarithmic correction appears (see~\cite{Bueno:2019mex} for example). 
{For (non-)Fermi liquid theories~\cite{Ogawa:2011bz}, another logarithmic violation to the area law is known.} For nonlocal~\cite{Shiba:2013jja} or non-translationally invariant~\cite{Vitagliano:2010db,Ram_rez_2014} systems, the volume law instead of the area law of EE has been confirmed. {To see the transition from the area law to the volume law, Lifshitz theories~\cite{He:2017wla,MohammadiMozaffar:2017nri,Gentle:2017ywk} might be an interesting playground as it possesses nonlocal feature in some limit.}

\section{Proof of the EE formulae of the vertex contributions}
\label{appencomp}
In the body of the paper, we have used the general formula for the vertex contributions to EE, such as in Eq.(\ref{EE-generalform2}). 
In this appendix, we prove that this formula gives all the contributions of a single vertex twist. 
In the case of the propagator contributions, the general formula is given by Eq.(\ref{e:EE2PIprop}) 
and the proof that all the single twist contributions are summarized by  the 1-loop expression is given 
in the 2PI framework in the previous papers \cite{Iso:2021vrk,Iso:2021rop}. 
For the vertex contributions, when auxiliary fields can be introduced, the proof is same, however, 
in general cases when various channels in the opened vertices are mixed, we need a different proof.
In this appendix, we give a diagrammatic proof. 

First, let us remind of the redundancies of assigning the flux $m$ of the plaquette 
to a twist of the propagators in the 1-loop diagram in Fig.\ref{Fig2}i. In this case, due to the $\mathbb{Z}_M$ gauge invariance at each vertex 
connecting propagators, the flux can twist only one of the propagators; not more than one, 
and this gives the coefficient $1/n$  in the expansion of Eq.(\ref{e:EE2PIprop}). 
The same happens for the vertex contributions. 
The configurations illustrated in Fig.\ref{f:verttwist} are interpreted as the  
vertex contributions to EE, but  a similar redundancy will occur when 
the corresponding composite operators form a 1-loop type diagram. 
Thus, in order for the proof, we will take the following two steps: 
 (i) summing all the vertex contributions as if all of them are independent
 and  then, (ii) taking account of the redundancies to obtain the correct vertex contributions. 
 This two-step proof  shows that only the 1-loop type contributions in Eq.(\ref{EE-generalform2}) survive. 
The proof is similar to the one based on the 2PI formalism.

Let us begin with the fundamental relation between the free energy and an $n$-point vertex. 
Suppose that we have an $n$-point interaction vertex whose action is given by
\ga{
(\text{action})=\frac{1}{2}\int d^{d+1}x\,\phi G_0^{-1}\phi+\cdots+
\frac{\lambda_n}{n} \int \prod_{i=1}^n d^{d+1}x_i 
\, V_{n0}(x_1,\cdots,x_n)\phi(x_1)\cdots\phi(x_n),\\
V_{n0}(x_1,\cdots,x_n)=\int d^{d+1}y\,\prod_{i=1}^{n}\delta^{d+1}(y-x_{i}).
}
Then, we have the equation
\aln{
\frac{\delta F}{\delta V_{n0}(x_1,\cdots,x_n)}=\frac{\lambda_n}{n}\langle\phi(x_1)\cdots\phi(x_n)\rangle ,
\label{nptfunc} 
} 
where  $F$ is the free energy and the right-hand side is the exact $n$-point function multiplied by the coupling constant. 

\begin{figure}[t]
\centering
\includegraphics[width=15cm]{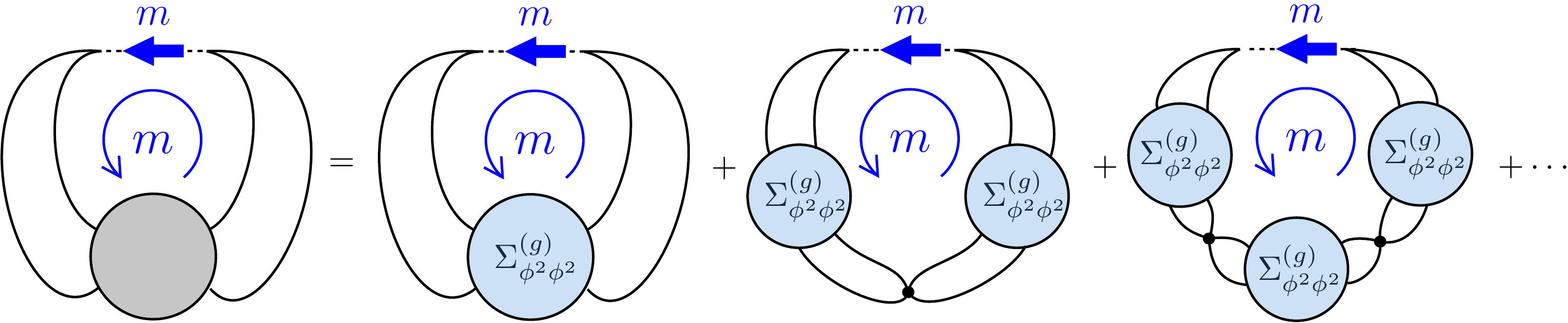}
\caption{Graphical expression of Eqs.(\ref{EEvertnaivek}) and (\ref{compexpand}) for $k=2$  in the $\phi^4$-theory. 
The gray blob on the left-hand side is the exact four-point correlation function. The dotted line denotes a
 twisted delta function to open the vertex. The twist of the diagram is given by the flux $m$ in the center circle. 
On the right-hand side, the twist is made associated with the propagator of a composite operator in the opened vertex. 
If we {would} open all the vertices on the circled line and take all the contributions to EE, 
it would give an overcounting of EE. 
}
\label{compredundancy}
\end{figure}

In order to evaluate the EE contributions from twisting vertices, let us first sum all the contributions
as if  they were  independent.  This can be done by taking a variation 
of bubble diagrams (free energy) with respect to the tree-level interaction vertex, 
and then reconnecting the endpoints by a set of free propagators as in the leftmost figure in Fig.\ref{compredundancy}.

According to Eq.(\ref{nptfunc}), 
if we naively summed all the contributions to EE from 
 opening all the $\lambda_n$-vertices, EE would be given by
\aln
{
S_{\lambda_n}^{\text{(naive)}}=&-\int d^{d+1}x\,d^{d+1}y\,\partial_M\left(\sum_{m=1}^{M-1}\delta^{d+1}(\hat{g}^m x-y)\right)\Bigg|_{M\to1}\,\nonumber\\
&\times\frac{\lambda_n}{n}\frac{1}{2}\Bigl\{C^{(n)}_{2,n-2}\,\langle\,[\phi^2](x)\,[\phi^{n-2}](y)\,\rangle+C^{(n)}_{3,n-3}\,\langle\,[\phi^3](x)\,[\phi^{n-3}](y)\,\rangle+\cdots\Bigr\}\nonumber\\
=&-\frac{V_{d-1}}{12}\int\frac{d^{d-1}k_\parallel}{(2\pi)^{d-1}}\frac{\lambda_n}{n}\Bigl\{C^{(n)}_{2,n-2}\,G_{\phi^2\phi^{n-2}}(\bm{0};k_\parallel)+C^{(n)}_{3,n-3}\,G_{\phi^3\phi^{n-3}}(\bm{0};k_\parallel)+\cdots\Bigr\},
\label{EEvertnaive}
}
where 
$C^{(n)}_{2,n-2}$, $C^{(n)}_{3,n-3}$, $\cdots$ are combinatorial factors 
to reconnect the endpoints. Endpoints can be decomposed into two sets as in Fig.\ref{compredundancy}
and then regarded as a composite operator. 
In the definition of $C^{(n)}_{2,n-2}$ etc., 
 we distinguish the left and right sets of  endpoints, $x$ and $y$, for later convenience, 
 and thus divide by 2 in the second line of Eq.(\ref{EEvertnaive})
 to avoid an overcounting. 

Here note that, 
we should not include a decomposition of  $n$ endpoints $\phi^n$ into $(\phi,\phi^{n-1})$ 
 because it does not corresponds to opening a vertex, rather it generates a non-1PI diagram in the 
 ordinary sense. 
 Such contributions would  lead to an overcounting of the propagator contributions. 
 Also, note  that we should not consider a reconnection of $n$ endpoints 
 in which some of them do not participate in the propagation of the composite operator;
 e.g. a diagram such that a pair of endpoints forms a closed loop and the other $n-2$ endpoints are
 decomposed into two sets to form the propagator of the composite operator.  
 This kind of diagrams 
 are  absorbed into the renormalizations of the coupling constant $\lambda_{n-2}$. 

Now let us go to step 2 to obtain the correct vertex contributions. 
$S_{\lambda_n}^{(\text{naive})}$ is not the correct one because of the redundancies we neglected.  
Let us consider the effects of redundancies  separately for each type of composite operator  in Eq.(\ref{EEvertnaive}). 

For simplicity, let us consider the $G_{\phi^k\phi^k}$-type contribution in Eq.(\ref{EEvertnaive})
which  emerges from a $2k$-point vertex by decomposing into 2 sets of $k$ and $k$. 
We simply  assume it is not mixed with other operators here. 
The simplest example  is $[\phi^2]$ in the $\phi^4$-theory, as described in Section \ref{s:vertexEE}. 
If the redundancies were neglected, 
the contribution to EE from this operator naively would take the form
\aln{
\left.S_{\lambda_{2k}}^{\text{(naive)}}\right|_{\phi^k\phi^k}
=-\frac{V_{d-1}}{12}\int\frac{d^{d-1}k_\parallel}{(2\pi)^{d-1}}\,\frac{\lambda_{2k}}{2k}\,C^{(2k)}_{k,k}\,G_{\phi^k\phi^k}(\bm{0};k_\parallel). 
\label{EEvertnaivek}
}
Reflecting the twisting of the composite operator, 
the Green function is restricted to $\bm{k}=\bm{0}$ modes. 
The Green function can be expanded with respect to the g-1PI self-energy {introduced in Sec.\ref{s:vertexEE}} as\footnote{
For a consistent expansion, we have defined the combinatorics factors ($C$'s) by distinguishing the two endpoints.}
\aln{
G_{\phi^k\phi^k}=\Sigma^{(g)}_{\phi^k\phi^k}&+\Sigma^{(g)}_{\phi^k\phi^k}\left(-\frac{\lambda_{2k}}{2k}C^{(2k)}_{k,k}\right)\Sigma^{(g)}_{\phi^k\phi^k}\nonumber\\
&+\Sigma^{(g)}_{\phi^k\phi^k}\left(-\frac{\lambda_{2k}}{2k}C^{(2k)}_{k,k}\right)\Sigma^{(g)}_{\phi^k\phi^k}\left(-\frac{\lambda_{2k}}{2k}C^{(2k)}_{k,k}\right)\Sigma^{(g)}_{\phi^k\phi^k}+\cdots.
\label{compexpand}
}
The correct formula must take the redundancies caused by $\mathbb{Z}_M$ gauge invariance
into account. Such redundancies occur in the above expansion of Eq.(\ref{compexpand}) when there are 
more than one $\Sigma^{(g)}_{\phi^k\phi^k}$ as shown in  Fig.\ref{compredundancy}. 
The coefficients of these terms in Eq.(\ref{compexpand}) overcount the effects of the twist. 

The resolution to avoid the overcounting   is simple. 
For the term consisting of $m$ g-1PI parts in Eq.(\ref{EEvertnaive}), we should divide it by $m$. 
Consequently, 
by replacing $G_{\phi^k\phi^k}$ in the naive estimation Eq.(\ref{EEvertnaivek}) with
\aln{
\Sigma^{(g)}_{\phi^k\phi^k}&+\frac{1}{2}\,\Sigma^{(g)}_{\phi^k\phi^k}\left(-\frac{\lambda_{2k}}{2k}C^{(2k)}_{k,k}\right)\Sigma^{(g)}_{\phi^k\phi^k}\nonumber\\
&+\frac{1}{3}\,\Sigma^{(g)}_{\phi^k\phi^k}\left(-\frac{\lambda_{2k}}{2k}C^{(2k)}_{k,k}\right)\Sigma^{(g)}_{\phi^k\phi^k}\left(-\frac{\lambda_{2k}}{2k}C^{(2k)}_{k,k}\right)\Sigma^{(g)}_{\phi^k\phi^k}+\cdots,
\label{complogexpand}
}
we get the correct contributions to EE as 
\aln{
\left.S_{\lambda_{2k}}\right|_{\phi^k\phi^k}
=-\frac{V_{d-1}}{12}\int\frac{d^{d-1}k_\parallel}{(2\pi)^{d-1}}\,\ln (1-\left(-\frac{\lambda_{2k}}{2k}C^{(2k)}_{k,k}\right)\Sigma^{(g)}_{\phi^k\phi^k}). 
}
This is the result of Eq.(\ref{e:EEcomp2}). 
In the 2PI formalism, the result is interpreted that only the 1-loop diagram provides a single 
twist contributions of propagators and all the other diagrams {cancel} each other. 
In the above discussions, we did not separate diagrams into 1-loop and others, but 
instead used the very basic relation of Eq.(\ref{nptfunc}). 
Then, using the property of the $\mathbb{Z}_M$ redundancy, the logarithmic factor for the 1-loop diagram
naturally appears. 

The above discussion can be straightforwardly generalized to more general composite operators with operator mixings. 
 When we have a set of operators $\{\mathcal{O}_a\}$ 
 by opening vertices, we consider   g-1PI self-energies $\Sigma^{(g)}_{\mathcal{O}_a\mathcal{O}_b}$ 
  and a matrix generalization of the nodal structure of   $(\lambda_n/n)\times C^{(n)}_{ab}$. 
  It is also straightforward  when the fundamental fields are mixed with other operators; 
  it is sufficient to consider  $\hat{\Sigma}^{(g)}\hat{G}_0$ in the formulation. 
  As a result, we arrive at the unified form of Eq.(\ref{EE-generalform2}).

\begin{acknowledgments}
We thank  Masahiro Hotta, Jun Nishimura,
 Takuma Nishioka, Yoshiki Sato, Kengo Shimada, Sotaro Sugishita, Takao Suyama, Tadashi Takayanagi, and 
 Kazuya Yonekura for discussions. 
We also acknowledge the referee of PRD to the previous papers 
whose comments have stimulated the present 
investigations. 
We are supported in part by the Grant-in-Aid for Scientific research, No. 18H03708 (S.I.), No. 16H06490 (S.I.),  No. 20J00079 (K.S.)
and SOKENDAI. 
\end{acknowledgments}


\bibliographystyle{apsrev4-1}
\bibliography{EE}
\end{document}